\documentstyle[aps,prb,multicol]{revtex}
\title{Isotope Effect for the Penetration Depth in Superconductors}
\author{A.~Bill, and V.Z.~Kresin}
\address{Lawrence Berkeley Laboratory, University of California,
Berkeley, CA 94720, USA}
\author{S.A.~Wolf}
\address{Naval Research Laboratory, Washington D.C.~20375-5343} 

\begin{document}
\maketitle
\begin{abstract}
We show that various factors can lead to an isotopic dependence of
the penetration depth $\delta$. Non-adiabaticity (Jahn-Teller
crossing) leads to the isotope effect of the charge carrier
concentration $n$ and, consequently, of $\delta$ in doped
superconductors such as the cuprates. A general equation relating the
isotope coefficients of $T_c$ and of $\delta$ is presented for
London superconductors. We further show that the presence of
magnetic impurities or a proximity contact also lead to an
isotopic dependence of $\delta$; the isotope coefficient turns out
to be temperature dependent, $\beta(T)$, in these cases.
The existence of the isotope effect for the penetration depth is
predicted for conventional as well as for high-temperature
superconductors. Various experiments are proposed and/or discussed.
\end{abstract}


\begin{multicols}{2}

\section{INTRODUCTION}\label{sec:intr}

When speaking of the isotope effect (IE) in superconductors, one
generally considers the influence of
isotopic substitution on the superconducting critical temperature
$T_c$. In systems where the pairing mechanism is at least partly
mediated by the electron-phonon interaction $T_c$ depends on phonon
energies. Thus, replacing one element by its isotope will affect
$T_c$ via the change in phonon frequency. In a series of
recent papers\cite{KW,KBWO,BK} we have shown that besides this
most simple case, there are a number of other factors, not
related to the pairing mechanism, that can
strongly affect the isotope coefficient
(IC) of $T_c$. The influence of the proximity effect, the presence of
magnetic impurities and non-adiabaticity were specifically studied.
Several experiments were proposed for conventional superconductors
and the oxygen IE in high-$T_c$ materials was discussed.
In the latter case it was experimentally shown (see
e.g.~Refs.~\onlinecite{franck,soerensen,zech,morris,zhao1} and
references therein), that the value of the isotope coefficient of
$T_c$ ranges from almost zero to values above $0.5$, depending on
the doping process.
The theory developed in Ref.~\onlinecite{KBWO} allowed us to
present a systematic description of all experimental data collected
to this day on the oxygen isotope effect of $T_c$ in
YBa$_2$Cu$_3$O$_{7-\delta}$ related systems. We stress here the fact,
however, that the theory is not restricted to high-$T_c$ materials,
but applies to conventional superconductors as well.

The influence on the IE of factors not related to the pairing
mechanism\cite{KW,KBWO,BK} raises the question, whether besides
$T_c$, there are other superconducting properties
that also display these new effects. In the present paper we focus
on one of the fundamental parameters, the
penetration depth, and demonstrate that, indeed, it can display a
non-trivial dependence on isotopic substitution. The treatment is
based on our previous analysis in Ref.~\onlinecite{KBWO} that
focused on the unconventional isotopic dependence of $T_c$.

After isotope substitution in a superconductor
($M \rightarrow M^{\star} = M + \Delta M$, $M$ is the ionic mass),
the value of the critical temperature is shifted:
$T_{c0}^{\star} = T_{c0} + \Delta T_{c0}$. As is known, the isotope
coefficient $\alpha_0$ is defined by the relation
$T_{c0}\sim M^{-\alpha_0}$ and is equal to
$\alpha_0 = - (M/\Delta M)(\Delta T_{c0}/T_{c0})$.
Suppose now that the superconductor is perturbed
by some external factor as e.g.~the presence of magnetic impurities
or a normal film (proximity effect). Superconducting
properties are affected by these factors.
For example, $T_{c0} \rightarrow T_c$ and, as a consequence, the
dependency of the new critical temperature on the ionic mass
($T_c \sim M^{-\alpha}$) will also be changed ($\alpha \ne \alpha_0$,
see Refs.~\onlinecite{KBWO,BK}). This not only affects the isotope
effect of $T_c$ but also, e.g., the IE of the penetration depth.

In the following, we are concerned with the isotopic dependence of
the penetration depth. By analogy with $T_c$, we define the new
isotope coefficient $\beta$ by the relation $\delta \sim M^{-\beta}$.
Therefore, $\beta$ is determined by
\begin{equation}\label{beta}
\beta = -  \frac{M}{\Delta M}\frac{\Delta\delta}{\delta} \qquad,
\end{equation}
where $\Delta\delta$ is the shift of the penetration depth induced
by isotopic substitution.

The structure of the paper is as follows. Section \ref{sec:NA} is
concerned with the non-adiabatic IE, whereas Secs.~\ref{sec:MI} and
\ref{sec:prox} address the impact of magnetic scattering and the
proximity effect on the IC, respectively. An interesting property
of the two latter factors affecting the IE is that the isotope
coefficient appears
to be {\it temperature dependent}. We not only demonstrate here that
such effects exist but that they are also sizeable. Moreover, we
argue that the measure of the IC can be used to probe the presence
of non-adiabaticity, magnetic impurities or the proximity effect in
superconductors. The theory leads us to propose several experiments
for conventional and high-temperature superconductors. We also
discuss how recent experiments done on
high-$T_c$ superconductors\cite{zhao1,zhao2} can be described by
our theory.

\section{ISOTOPE EFFECT OF $\delta$ AND NON-ADIABATICITY}
\label{sec:NA}
The theory for the non-adiabatic IE of $T_c$ has recently been
proposed\cite{KW} and applied to describe the oxygen IC in
high-temperature superconductors\cite{KBWO,BK} as well as properties
of manganites\cite{KW2}. We show in this section that other
quantities like the penetration depth $\delta$ of a magnetic
field can be affected by isotopic substitution through this
non-adiabatic channel.

\subsection{Theory}
\label{subsec:theoryNA}
\subsubsection{Isotope effect of $n$}
\label{subsubsec:IEns}
Let us describe the main concept of the non-adiabatic IE introduced
in Ref.~\onlinecite{KW} (see also Ref.~\onlinecite{KBWO}).
Consider a system composed of a conducting system and
a charge reservoir (as for example the CuO$_2$ plane and the buffer
layer of a cuprate high-$T_c$ superconductor) and assume that
there is a charge transfer between these subsystems. In general, the
reservoir--conducting-layer charge-transfer process occurs through a
group of atoms located in both subsystems and/or bridging the two.
For example, the charge transfer between the CuO$_2$ plane and the
reservoir (CuO chain for YBCO, BiO plane for Bi-based oxides, etc...)
occurs through the apical oxygen. The electronic degeneracy and the
corresponding Jahn-Teller instability of the group of atoms leads to
the crossing of electronic terms.
Because of the breakdown of the adiabatic
approximation for these atoms, the
electronic and ionic degrees of freedom can no longer be considered
as decoupled.
The potential energy surface considered as a function of the ionic
coordinate
is composed of two electronic terms with two close minima.
For example, the motion of the apical oxygen which is a lighter
ion of the Cu-O-Cu Jahn-Teller complex is characterized by two minima
(a more detailed discussion is given in Refs.~\onlinecite{KW,KBWO};
experimental evidence of these minima can be found in
Refs.~\onlinecite{mustre,sharma}).
Note that each of these terms can be considered in the harmonic
approximation. The ``double-well'' is thus not related to the
anharmonicity of the lattice, but to the non-adiabatic behaviour of
the group of ions involved in the charge-transfer process. In a
qualitative picture, the charge
transfer between the conducting and the reservoir subsystems will
involve the tunneling of the ionic complex between the two minima
and, as a consequence, the density of charge carriers in the
conducting subsystem will depend on the ionic masses. Properties
depending on the charge-carrier density $n$ will thus be ionic mass
dependent ($n\equiv n(M)$) and exhibit an isotope effect.

In order to consider the Jahn-Teller crossing (where the adiabatic
approximation is not applicable) it is convenient to use the
so-called diabatic representation (see, e.g.,
Ref.~\onlinecite{omalley}). In this representation one can
show\cite{KBWO} that the energy level splitting has the form
\begin{eqnarray}\label{H12}
H_{12} = < \Psi_1 | H_e | \Psi_2 > \simeq L_0 F_{12} \qquad ,
\end{eqnarray}
with
\begin{eqnarray}
L_0 &=& \int \text{d{\bf r}}
\psi^{\ast}_1({\bf r,R_0}) H_e \psi_2({\bf r,R_0}) \qquad ,
\nonumber\\
\label{FC}
F_{12} &=& \int \text{d{\bf R}}
\Phi^{\ast}_1({\bf R})\Phi_2({\bf R}) = F_{12}(M) \qquad .
\end{eqnarray}
$H_e$ is the electronic part of the Hamiltonian and
$\Psi_j({\bf r,R}) = \psi_j({\bf r,R})\Phi_j({\bf R})$ ($j=1,2$).
$\psi_j({\bf r,R})$ is the electronic wave function (with electronic
coordinates {\bf r}) depending parametrically on the nuclear
coordinates {\bf R} and $\Phi_j({\bf R})$ is the vibrational wave
functions. The last equality in Eq.~(\ref{H12}) was obtained
because $\psi_j$ ($j=1,2$) is a slow varying function of
{\bf R} and can thus be evaluated at ${\bf R}_0$, the crossing
of electronic terms. $L_0$ does not depend on the ionic masses. On
the other hand, the important Franck-Condon factor $F_{12}$ depends
on the lattice wave functions $\Phi_j$ and thus on ionic masses $M$.

The reservoir--conducting-layer charge transfer is accompanied by a
change of electronic terms and can be visualized
as a multi-step process. First,
the charge carrier can move from the reservoir to the
group of ions (e.g.~from the chains to the apical oxygen in YBCO).
Then, the complex tunnels to the other electronic term
($\Psi_1\rightarrow\Psi_2$). As a final step, the charge carrier can
hop to the conducting layer.
We emphasize that this charge transfer is a dynamical process and
leads to a stationary state in the sense that the charges oscillate
in time between the reservoir and the conducting subsystems, with
a frequency given by Eq.~(\ref{H12}). Since the ionic masses
affect $F_{12}$, and thus $H_{12}$, different isotopic masses imply
different values of the characteristic charge-transfer frequency.
This in turn affects the charge-carrier concentration $n$.

The isotope coefficient of $T_c$ can be written as
$\alpha = \alpha_{ph} + \alpha_{na}$, where
$\alpha_{ph} = (M/T_c)(\partial T_c/
\partial \Omega)(\partial \Omega/\partial M)$ is the usual (BCS)
phonon contribution ($\Omega$ is a characteristic phonon energy)
and the non-adiabatic contribution is given by:
\begin{eqnarray}\label{TcNA}
\alpha_{na} =
\gamma \frac{n}{T_c}\frac{\partial T_c}{\partial n} \qquad ,
\end{eqnarray}
where the parameter $\gamma = - M/n (\partial n/\partial M)$ has a
weak logarithmic dependence on $M$ (see Ref.~\onlinecite{KW}).
Eq.~(\ref{TcNA}) shows that the IC of $T_c$ depends on the
doping of the conducting layer and on the relation $T_c(n)$. This
result was used in Ref.~\onlinecite{KBWO} to analyse the IE of
high-temperature superconductors.

\subsubsection{Non-adiabatic isotope effect of $\delta$}
\label{subsubsec:IEdelta}
Let us now focus on the non-adiabatic IE of the penetration depth.
As shown above (see also Refs.~\onlinecite{KW,KBWO}),
non-adiabaticity for doped materials such as cuprates, leads to the
dependence of the charge-carrier concentration $n$ on the ionic
mass $M$, that is $n = n(M)$.
In the weak-coupling London limit the penetration depth is given by
the well-known relation
\begin{equation}\label{london}
\delta^{2} = \frac{m c^2}{4\pi n_s e^2}
= \frac{m c^2}{4\pi n\varphi(T/T_c) e^2} \qquad .
\end{equation}
where $m$ is the effective mass. $n_s$ is the superconducting
density of charge carriers, related to the normal density $n$
through $n_s = n\varphi(T/T_c)$. The function $\varphi(T/T_c)$ is a
universal function of $T/T_c$. For example, for conventional
superconductors,
$\varphi \simeq 1-(T/T_c)^4$ near $T_c$, whereas $\varphi \simeq 1$
near $T=0$ (in the absence of magnetic impurities; their
influence is discussed at the end of this and in the following
sections). Note that the relation $\delta^{-2}\sim n_s$ is also
valid in the strong-coupling case (see, e.g., Refs.~\onlinecite{BK}).

We can now determine the isotope coefficient $\beta$ of the
penetration depth from the relation (\ref{london}):
\begin{eqnarray}\label{betans}
\beta \equiv
-\frac{M}{\delta}\frac{\partial\delta}{\partial n_s}
\frac{\partial n_s}{\partial M}
= \frac{M}{2 n_s}\frac{\partial n_s}{\partial M} \qquad .
\end{eqnarray}
Because of the relation $n_s = n\varphi(T)$, one has to
distinguish two contributions to $\beta$. There is a usual (BCS)
contribution, $\beta_{ph}$,
arising from the fact that $\varphi(T/T_c)$ depends on ionic mass
through the dependency of $T_c$ on the characteristic phonon
frequency. Indeed, isotopic substitution leads to a
shift in $T_c$ and thus in $\delta$, which might be noticeable near
$T_c$ (see the discussion in Sec.~\ref{subsec:discNA}). In this
paper, however, we focus on the non-trivial manifestation of
isotopic substitution, arising from the isotope dependence of the
charge-carrier concentration $n$.

From Eq.~(\ref{betans})  and the relation $n_s = n\varphi(T)$ it
follows that
\begin{eqnarray}\label{betaNA}
\beta = \beta_{na} + \beta_{ph}
\end{eqnarray}
where
\begin{eqnarray}\label{betaNA1}
\beta_{na} = \frac{M}{2n}\frac{\partial n}{\partial M}
\qquad ,\\
\label{betaNA2}
\beta_{ph} =
\frac{M}{2\varphi(T)}\frac{\partial\varphi(T)}{\partial M} \quad .
\end{eqnarray}
Note that $n(M)$ is the {\it normal}-state charge-carrier
concentration. Thus, unlike $\beta_{ph}$ (which depends on
$\Phi(T/T_c)$), the non-adiabatic coefficient $\beta_{na}$ does
{\it not} depend on parameters characterizing the superconducting
state. This effect should be observed near $T=0$K, where $\beta_{ph}$
is negligibly small.\\
Comparing Eq.~(\ref{TcNA}) and Eq.~(\ref{betaNA1}) one infers that
$\beta_{na} = -\gamma/2$ and
thus establish a relation between the non-adiabatic isotope
coefficients of $T_c$ and $\delta$:
\begin{eqnarray}\label{ICTcNA}
\alpha_{na}
= -2\beta_{na} \frac{n}{T_c}\frac{\partial T_c}{\partial n}
\qquad .
\end{eqnarray}
This result holds for London superconductors. The equation contains
only measurable quantities and can thus be verified experimentally.
 It is interesting to note that
$\beta_{na}$ and $\alpha_{na}$ have opposite signs when
$\partial T_c/\partial n > 0$ (which corresponds to the underdoped
region in high-$T_c$ materials).

In later sections we will discuss the influence of magnetic
impurities on the isotope effect of $T_c$ and $\delta$. We note
here that in the presence of magnetic impurities the
relation $n_s = n\varphi(T)$ remains valid, but $\varphi(T)$ depends
now on the direct scattering amplitude
$\Gamma_2$ defined in Sec.~\ref{subsec:MIT0} (this results, e.g., in
the inequality $n_s(T=0) < n$ in the gapless regime). As a
consequence, magnetic impurities affect $\beta_{ph}$ in
Eq.~(\ref{betaNA2}), but leave $\beta_{na}$ and thus
Eq.~(\ref{ICTcNA}) unchanged.

\subsection{Non-adiabatic isotope effect in high-$T_c$ superconductors}
\label{subsec:discNA}
The value of the parameter $\gamma$ (see Eq.~(\ref{TcNA})) has been
evaluated in Ref.~\onlinecite{KBWO} for Pr-substituted (YPrBCO) and
oxygen-depleted (YBCO) YBa$_2$Cu$_3$O$_{7-\delta}$ and was found to
be $\gamma = 0.16$ and $\gamma = 0.28$ respectively. With
Eq.~(\ref{ICTcNA}) one then obtains the IC of the penetration
depth $\beta_{na} = -0.08$ in the first case and
$\beta_{na} = -0.14$ for the latter.

Experiments have been performed on YBCO\cite{zhao1} and
La$_{2-x}$Sr$_x$CuO$_4$\cite{zhao2} (LSCO) which indicate that the
Meissner fraction (and thus the penetration depth) indeed displays an
isotope shift. Unfortunately, the measurements on YBCO have been
done near $T_c$ where {\it any} superconductor
(conventional or high-$T_c$) displays an isotope shift of $\delta$
through the dependency $\delta(T/T_c) \sim \sqrt{1 - T^4/T_c^4}$
 and corresponds thus to the usual BCS IE (this is the
contribution $\beta_{ph}$ arising through $\varphi(T)$ as discussed
in the last section; see also
Ref.~\onlinecite{lynton}). The BCS isotope coefficient of $\delta$
(near $T_c$) can easily reach values of the order observed in the
experiment\cite{zhao1}, even for a very small IC of $T_c$
(with a value $\alpha = 0.025$ as observed in
YBa$_2$Cu$_3$O$_{7-\delta}$ one obtains $\beta\sim -0.6$ for
$T/T_c \sim 0.95$). To avoid the BCS contribution to the IC that
appears in all superconductors because of the strong dependency
$\delta(T/T_c)$ near $T_c$ it is better to study experimentally the
IE of the
penetration depth near $T=0$. This statement is valid if one is
interested in determining the non-adiabatic contribution to the IC
$\beta$ but is not general. As will be shown below, it is for example
possible to extract the influence of magnetic impurities on the IE
even near $T_c$.

The situation with the measurements of Ref.~\onlinecite{zhao2} is
different. The experiments\cite{zhao2} done on LSCO were obtained
near $T=0$ and reflect the unconventional dependence $\delta(M)$
(the BCS contribution to the IC of $\delta$ vanishes at these
temperatures). The shift
has been measured near $x\simeq 0.11$ and $x\simeq 0.15$.

Near the
concentration $x\simeq 0.11$, $T_c$ experiences a small depression
as a function of doping $x$.\cite{crawford}
Although several explanations have been proposed (as, e.g.,
the presence of electronic inhomogeneities, a change
of the electronic density of states due to lattice distortions,
impurity scattering, magnetic ordering, etc...\cite{dabrowski}) the
presence of the dip is not well understood yet.
The choice of this concentration is thus unfortunate and
inappropriate for the study of the IE.
It was argued\cite{zhao2} that there is no influence
of isotopic substitution on the charge carrier concentration $n$
near $x\simeq 0.11$ and that $n$ can therefore not depend on the
ionic mass. To support this idea the influence of isotopic
substitution on the structural (tetragonal to orthorhombic)
transition temperature $T_s$ was studied\cite{zhao2}. The
assertion is based on the assumption that $T_s$ depends solely on
$n$, that there is a monotonic relation between $T_s$ and $n$ with
non-zero slope and a one-to-one correspondance between $n$ and $x$.
However, the experiments performed on LSCO \cite{dabrowski}
show that the hole concentration $n$ dependency of $T_c$ and
$T_s$ as well as the relation $n(x)$ are not well established,
especially in the region of the dip. The structural phase transition
temperature $T_s$ has also only been measured at three points in the
vicinity of the dip\cite{crawford,dabrowski,zhao2}. Since $T_c$
has an unexpected behaviour it would be of interest
to study the detailed dependency $T_s(x)$ in this region.
Furthermore, as known from other high-temperature superconductors
(as, e.g., YBCO) it is unlikely that the relation between $n$ and
$x$ is linear. We emphasize that our analysis of the IC from
Eqs.~(\ref{TcNA}) and (\ref{ICTcNA}) does not require any assumption
on the relation $n(x)$ since we rely solely on the experimental
relation as obtained, e.g., from $\mu$SR experiments (in this case
one has $T_c(\sigma\sim n_s/m^{\star})$; see
Ref.~\onlinecite{BK}).
In short, since the structural and electronic properties
are not well understood in the region around $x\simeq 0.11$, it does
not allow a conclusive statement on the IE of $\delta$.

The only experimental result\cite{zhao2} that can be
discussed, and that does not contain the BCS contribution is the one
done at $x\simeq 0.15$ (for LSCO). Using
Eq.~(\ref{beta}) and the value of the experimentally observed
relative shift $\Delta\delta/\delta = 2\%$\cite{zhao2} one obtains
$\beta_{na} \simeq 0.16$ [and $\gamma = 0.32$, cf.~Eq.~(\ref{TcNA})
and (\ref{ICTcNA})] which is in good agreement with the
calculations presented above. Indeed, it is of the same order as
$\beta_{na}$ for
oxygen-depleted YBCO. This is consistent with the fact that in both
these materials the reservoir-CuO$_2$-plane charge transfer involves
the same group
of ions including the apical oxygen. The larger value for LSCO can be
traced back to the fact that in YBCO one has taken into consideration
the presence of magnetic impurities (which also enhances the IC),
whereas in LSCO there is no indication of the presence of significant
quantities of such impurities.\cite{zhao2}$^c$

One notes further that the IC of $\delta$ is much larger in YBCO and
LSCO than in YPrBCO. The reason for this discrepancy can be
twofold. On the one hand,
the charge-transfer channel is different in YPrBCO from the two
other materials\cite{KBWO} (e.g.~it does not involve the apical
oxygen).
On the other hand, the concentration of magnetic impurities is
higher in YPrBCO than in YBCO (and LSCO). This was taken into account
in the calculation of $\alpha_{na}$ and influences the value of
$\beta_{na}$ in YBCO related systems as well. A larger part of the
IC is thus due to the magnetic impurities in YPrBCO (see
Ref.~\onlinecite{KBWO} for a detailed study of the interplay between
non-adiabaticity and magnetic impurities).

One should add that the experiment determining the IC through
magnetic susceptibility measurement\cite{zhao1,zhao2} is rather
inaccurate. More reliable data could be obtained by measuring the
penetration depth shift using microwave measurements or the
Josephson effect.
It would be interesting to perform experiments on the systems
discussed above that display the non-adiabatic charge-transfer
channel.

As stressed earlier and according to the
analysis of Ref.~\onlinecite{KBWO} one has, however, to include the
contribution of magnetic impurities to describe correctly the IE of
$T_c$ in some of the high-$T_c$ superconductors. The magnetic
impurities also directly affect the IC of $\delta$ and should be
treated as well. This will be done in the next section.

\section{ISOTOPE EFFECT OF $\delta$ AND MAGNETIC IMPURITIES}
\label{sec:MI}

As described in Refs.~\onlinecite{KBWO,BK} magnetic impurities modify
the IC of $T_c$. We show here that it can also induce an IE of
the penetration depth. Consider a superconductor (conventional or
high$T_c$) doped with magnetic impurities. Abrikosov and
Gor'kov\cite{AG} have shown that these impurities act as pair
breakers. In the dirty limit one can calculate the penetration depth
analytically for temperatures near $T_c$ and at $T=0$.

\subsection{Magnetic impurity contribution to $\beta$ near $T_c$}
\label{subsec:MITc}
The penetration depth $\delta$ in the presence of magnetic impurities
and near $T_c$ has been calculated by Skalski
{\it et al.}\cite{skalski}. Taking terms up to the order
$\Delta^2$ the result is given by\cite{skalski}:
\begin{eqnarray}\label{delMI}
\delta^{-2} = \sigma \frac{\Delta^2(T)}{T_c}
\zeta(2,\gamma_s + \frac{1}{2}) \qquad ,
\end{eqnarray}
where $\sigma = 4\sigma_N/c$ ($\sigma_N$ is the normal state
conductivity), $\zeta(z,q) = \sum_{n\ge 0} 1/(n+q)^z$
and $\gamma_s = \Gamma_s/2\pi T_c$ ($\Gamma_s = \tilde{\Gamma}_s n_M$
is the spin-flip scattering amplitude; $n_M$ is the
concentration of magnetic impurities and $\tilde{\Gamma}_s$ is a
constant). Near $T_c$, the order parameter can be written
as\cite{skalski}:
\begin{eqnarray}\label{gapMI}
\Delta^2 = 2\Gamma_s^2(1-\tau)
\frac{1 - \bar{\zeta}_2 + (1-\tau)\left[ \frac{1}{2} -\bar{\zeta}_2 + \bar{\zeta}_3 \right]}{\bar{\zeta}_3 - \bar{\zeta}_4 }
\equiv 2\Gamma_s^2 \frac{N_1}{D_1}
\end{eqnarray}
with $\bar{\zeta}_z = \gamma_s^{z-1} \zeta(z,\gamma_s + 1/2)$,
$z=1,2,\ldots$ and $\tau = T/T_c$. Using Eqs.~(\ref{delMI}) and
(\ref{gapMI}) one can calculate the isotope coefficient of the
penetration depth in the presence of magnetic impurities
($\beta_{m}$).
In order to single out the impact of magnetic impurities on the
value of the IC, we calculate the quantity $\tilde{\beta}_{m} =
-(M/\Delta M)(\Delta\tilde{\delta}/\tilde{\delta})$
[cf.~Eq.~(\ref{beta})],
where $\tilde{\delta} = \delta(T,\Gamma_s)/\delta(T,0)$. As a result,
the IC is written as a difference $\tilde{\beta}_m = \beta_m - \beta_0$
where $\beta_m$ ($\beta_0$) is the isotope coefficient of
$\delta(T,\Gamma_s)$ ($\delta(T,0)$). Using the
relation $(\partial/\partial q) \zeta(z,q) = - z\zeta(z+1,q)$
one can express $\tilde{\beta}_m$ in the form:
\begin{eqnarray}\label{icMI}
\tilde{\beta}_{m} = (R_1 - R_0)\alpha_{m} \qquad ,
\end{eqnarray}
where
\begin{eqnarray}
\label{R1}
R_1 &=& -\frac{1}{2}f_1 = - \frac{1}{2}\left( \frac{N_2}{N_1} - \frac{D_2}{D_1}
+ 2\frac{\bar{\zeta}_3}{\bar{\zeta}_2} - 1 \right)  \qquad ,\\
\label{R0}
R_0 &=& - \frac{1}{2}\frac{\alpha_0}{\alpha_m}f_0
= -\frac{1}{2}\left[ 1 - \psi^{\prime}(\gamma_s + 1/2)\gamma_s \right]f_0
\qquad ,
\end{eqnarray}
and $f_0 = (3-\tau^2)/(1-\tau)(3-\tau)$.
The functions $N_1$, $D_1$ are defined in Eq.~(\ref{gapMI}) and
\begin{eqnarray*}
N_2 &=& 3(1-\tau)^2
\left[ \bar{\zeta}_2 - 2\bar{\zeta}_3 + \bar{\zeta}_4  \right]
+ \tau(2-\tau) - \bar{\zeta}_2\\
D_2 &=& 2\left(3\bar{\zeta}_4 - \bar{\zeta}_3 - 2\bar{\zeta}_5\right)
\qquad .
\end{eqnarray*}
Furthermore, Eq.~(\ref{R0}) has been written using the relation for the
isotope coefficient of $T_c$ in the presence of magnetic
impurities:\cite{KBWO,carbotte}
\begin{eqnarray}\label{ICTcMI}
\alpha_{m} =
\frac{\alpha_0}{1 - \psi^{\prime}(\gamma_s + 1/2)\gamma_s} \qquad ,
\end{eqnarray}
where $\alpha_0$ is the isotope coefficient of $T_{c0}$, the critical
temperature in the absence of magnetic impurities.
$\psi^{\prime}$ is the derivative of the psi function.

Eq.~(\ref{icMI}) shows that the isotope coefficient of $\delta$ is
proportional to the isotope coefficient of $T_c$. This relation is
valid near $T_c$ (where $\Delta$ is small) and for
impurity concentrations such that the condition
$\Delta T_c/T_c \ll 1$ is satisfied.

An important feature of the result (\ref{icMI}) is that the IC of
$\delta$ is a universal function $\beta(T/T_c)$.
Indeed, Eq.~(\ref{icMI}) can be calculated once $T/T_c$, $T_{c0}$,
$\alpha_0$ and $\gamma_s$ are given. The three first quantities are
obtained from experiment. The last parameter $\gamma_s$
also determines the depression of
$T_c$ induced by magnetic impurities and can be calculated from
the Abrikosov-Gor'kov equation\cite{AG}:
\begin{eqnarray}\label{TcAG}
\ln\left(\frac{T_{c0}}{T_c}\right)
= \psi\left(\frac{1}{2} + \gamma_s \right)
- \psi\left(\frac{1}{2}\right) \qquad .
\end{eqnarray}
From Eqs.~(\ref{icMI}), (\ref{ICTcMI}) and (\ref{TcAG}) we thus
conclude that the knowledge of the measurable quantities
$\alpha_0$, $T_{c0}$ and $T/T_c$
completely determines the IC $\tilde{\beta}_{m}$.

If besides the magnetic impurity channel there is
also a contribution from the non-adiabatic channel as presented in
the last section, the total IC takes the form $\beta_{tot} =
\beta_{ph} + \beta_{m} + \beta_{na}$. In addition, according to
Ref.~\onlinecite{KBWO}, one has to replace $\alpha_0$ in Eq.~(\ref{ICTcMI}) by
$\alpha_{na}+\alpha_{ph}$ ($\alpha_{na}$ is given in Eq.~(\ref{TcNA})
and $\alpha_{ph}$ is the usual phonon contribution to the IC of
$T_c$). The non-adiabatic and magnetic-impurity contributions to
the IC are thus non-additive near $T_c$ (because of the
presence of $\alpha_{na}$ in $\beta_{m}$ and the presence of
$\Gamma_2$ in $\beta_{ph}$; as shown in Sec.~\ref{subsubsec:IEdelta},
$\beta_{na}$ is unaffected by magnetic impurities). One notes
further, that
if the IC is induced by magnetic impurities only, then its value
is always negative, whereas the situation might be different in
the presence of a non-adiabatic contribution.

The factor $\beta_0$ in Eq.~(\ref{icMI}) which is the IC of $\delta$
in the absence of magnetic impurities contains a trivial
temperature dependence due to the factor $\varphi(T/T_c) \sim
1 - (T/T_c)^4$ and the isotopic dependence of $T_c$. However, the
presence of magnetic impurities leads to an additional, novel
{\it temperature} dependence of $\tilde{\beta}_{m}$, beyond the BCS
one. This
temperature dependence is shown in Fig.~1 for fixed values of $T_c$
(i.e.~of the magnetic impurity concentration). The
solid and dashed lines describe the case where only magnetic
impurities are added to the system (no non-adiabatic contribution).
The situation corresponds to Zn-doped YBCO (YBCZnO). It has been
shown with use of NMR technique that Zn-substitution (for Cu)
occurs in the CuO$_2$ plane and leads to the appearence of localized
magnetic moments (see Ref.~\onlinecite{zagoulev} and below). The two
other
curves (dotted and dash-dotted lines) are obtained when, in addition,
a non-adiabatic channel is present. They are
discussed below. Fig.~1 shows that the absolute value of the IC
increases as one approaches $T_c$. This temperature dependency is
significant since in certain cases the value of $\beta$ triples when
$\tau = 0.75$ increases to $\tau = 0.95$. Note also, that
the solid and dashed lines describe conventional superconductors
as well, since only $\alpha_0$ has to be modified (in YBCZnO
$\alpha_0 = \alpha_{ph} = 0.025$ which is much smaller than
typical values for conventional superconductors).

Fig.~2 shows the dependency of the IC $\tilde{\beta}_{m}$ on $T_c$
(i.e.~on the concentration $n_M$) for fixed values of the
temperature. It appears that an increase of magnetic impurity
concentration leads to an increase of $|\tilde{\beta}_{m}|$.
Moreover, the change of $\tilde{\beta}_m(T_c)$ is stronger as $T$
approaches $T_c$. It turns out that the qualitative behaviour of
$\tilde{\beta}_m(T_c)$ is
similar to $\alpha_{m}(T_c)$ [Eq.~(\ref{ICTcMI}) with
$\alpha_0 = \alpha_{ph}$] but with opposite sign (see also Fig.~1 of
Ref.~\onlinecite{KBWO}). As for Fig.~1, the parameter $\alpha_0$
corresponds to the value for YBCZnO, but the qualitative picture
holds for conventional superconductors as well.

The calculation of the IC near $T_c$ presented above suggests an
experiment to determine the contribution of magnetic impurities
to the IC. For conventional superconductors where the magnetic
moments are introduced into the superconductor as impurities one
can substract the BCS isotope coefficient discussed at the end of
last section by measuring its temperature dependency.
For high-temperature superconductors the same procedure can in
principle be applied. Nevertheless, one has to be more carefull
since magnetic impurities are intrinsically present in some of these
materials as shown in Ref.~\onlinecite{furrer}. Indeed, there
are evidences that already optimally doped YBCO contains magnetic
impurities. Upon depletion of oxygen one further
adds magnetic impurities. One can thus not avoid the presence of
these localized magnetic moments. As a consequence, it is more
difficult to extract the BCS component for these materials. The
procedure in high-$T_c$ superconductors
would thus consist in measuring the (temperature dependent)
isotope coefficient of $\delta$ at optimal doping (which contains
the smallest concentration of intrinsic magnetic moments) and
substract this component
from the measurements at a given doping. This applies, e.g., to
oxygen-depleted, Pr or Zn-doped YBCO (see also
Refs.~\onlinecite{zagoulev,furrer}).

Note that our analysis in
this section is based on the theory in Ref.~\onlinecite{AG} (see
Eq.~(\ref{TcAG})). The core of this theory is the introduction of the
pair-breaking effect. This effect requires singlet pairing and is
applicable to both $s$ and $d$-wave symmetries of the order
parameter. There is, of course, a quantitative difference between
these two cases, but the isotope effect for the penetration depth
should be observed, regardless of the symmetry.
In this paper, in analogy with Ref.~\onlinecite{KBWO}, we make
specific calculations for the $s$-wave scenario. Therefore, these
calculations are valid, even quantitatively for conventional
superconductors, as well as for some oxides (e.g.~Nd-based cuprates,
Ba-Pb-Bi-O). As for the YBCO compound, our calculations of the
isotope shift for $T_c$ caused also by magnetic impurities, revealed
excellent agreement with the data (see Ref.~\onlinecite{KBWO},
Fig.~1a) which is interesting from the point of view of analyzing
the complex problem of the symmetry. In this paper, we employ a
similar approach. It is essential, however, to note that, as
mentioned above, the qualitative picture is similar for both
scenarios ($s$ or $d$-wave).

The results presented in Figs.~1 and 2 were calculated on the basis
of the parameters for Zn-doped YBCO (YBCZnO) and consider solely the
influence of magnetic impurities on $\beta$. The situation is more
complicated if, in addition, non-adiabaticity is present in the
system. In this case $\tilde{\beta}_m$ is given by Eqs.~(\ref{icMI})
to
(\ref{TcAG}) and $\alpha_0$ by Eq.~(\ref{TcNA}). To stress the
influence of non-adiabaticity on $\tilde{\beta}_m$ through $\alpha_0$
we write $\tilde{\beta}_{m+na}$ in the following. We have calculated
the IC of the penetration depth near $T_c$ for Pr-doped YBCO
(YPrBCO) using the parameters derived in Ref.~\onlinecite{KBWO}.
The results are shown in Figs.~1 (dotted and dot-dashed lines) and 3
for the temperature and magnetic impurity concentration
dependence, respectively.

Several comments can be made by analyzing these figures.
Fig.~1 shows that there is no qualitative change in $\beta(T)$ (at
fixed $T_c$) when adding the non-adiabatic contribution.
Non-adiabaticity merely induces a stronger temperature dependency.
This may, however, be specific to the materials studied here.
On the other hand, contrary to the case of magnetic impurities alone
(Fig.~2), there is a qualitative difference (besides the opposite
signs) between the IC of $T_c$ and $\delta$ taken as a function of
$T_c$ and in the presence of non-adiabaticity. To understand this,
one has to come back to Eqs.~(\ref{icMI})-(\ref{R0}). Let us first
rewrite Eq.~(\ref{icMI}) in the form
\begin{eqnarray}
\tilde{\beta}_{m} = \beta_m - \beta_0
= - \frac{1}{2}\left( f_1 \alpha_m - f_0 \alpha_0 \right) \qquad ,
\end{eqnarray}
where $f_0$ and $f_1$ are defined in Eqs.~(\ref{R1}),(\ref{R0}). It
turns out that $f_1(T/T_c)$ and $f_0(T/T_c)$ are never more than
several percents apart from each other for given $T$ and $T_c$. The
shape of $\tilde{\beta}_{m+na}$ in the presence of non-adiabaticity
and magnetic impurities is thus
mainly determined by the difference $\alpha_{m} - \alpha_0$ [where
$\alpha_{m}$ is given by Eq.~(\ref{ICTcMI}) and $\alpha_0 =
\alpha_{ph} + \alpha_{na}$ by
Eq.~(\ref{TcNA})]. Using a polynomial expansion of
$T_c(n)$ (as done in Ref.~\onlinecite{KBWO}) one can show that
$\alpha_0$ saturates with decreasing $T_c$
[$\alpha_{na}(T_c\rightarrow 0)\rightarrow \gamma$, with $\gamma$
defined in Eq.~(\ref{TcNA})], whereas $\alpha_{m}$ increases as
$n_M^2$ in the same limit (see Ref.~\onlinecite{KBWO}). Thus, at high
impurity doping, the behaviour of $\tilde{\beta}_{m+na}$ is dominated
by the denominator of Eq.~(\ref{ICTcMI}), that is by the magnetic
impurities. In the opposite low doping regime, 
$\tilde{\beta}_{m+na}$ is dominated by the non-adiabatic part in
$\alpha_{m}$ and $\alpha_0$.
Accordingly, $\tilde{\beta}_{m+na}$ behaves similarly near optimal $T_c$ and
near the 60K plateau (in YBCO), where $\partial T_c/\partial n$ is small. Thus,
as can be seen on Fig.~3b, $\tilde{\beta}_{m+na}$ behaves approximately as
$-\alpha_{na}$ in these regions but is smaller by an amount approximately
equivalent to $\alpha_0$. As one gets away from the plateaus, the
magnetic impurity part of $\alpha_{m}-\alpha_0$ becomes dominant,
leading to a departure from the behaviour of $\alpha_{na}$. The fact
that $\tilde{\beta}_{m+na}$ is mainly determined by
$\alpha_m - \alpha_0$ (instead of $\alpha_m$ for the IC of $T_c$) explains
why the change of curvature of $\tilde{\beta}_{m+na}$ (see Fig.~3b) takes
place at higher $T_c$'s (lower magnetic impurity concentrations)
than the IC of $T_c$, $\alpha_{m}$ (see also Fig.~2a of Ref.~\onlinecite{KBWO}).
The change of curvature should be measurable since the precision of
measurements increases with
decreasing doping\cite{soerensen}. Accordingly, the effect should
also be observable for oxygen-depleted YBCO for $T_c$ near its
optimal value and below but near the 60K plateau.

We have not calculated the penetration depth IC for oxygen-depleted
YBCO near $T_c$ because
the dependencies $\Gamma_s(x)$ (the spin-flip scattering amplitude
as a function of oxygen doping) and $n(x)$ are
not well known for this material. This was not a major problem
for the calculation of the IC $\alpha_{m}$ of $T_c$, because
$\alpha_{m}$ is
mainly determined by the non-adiabatic channel in the regime where the
experiments were performed and is affected by
magnetic impurities in the higher doping (i.e.~lower $T_c$, away
from the plateaus) regime (see Eq.~(\ref{ICTcMI})
and Ref.~\onlinecite{KBWO}). However, it might be of importance for
the calculation of the IC of $\delta$ because of the more complicated
structure of the relation $\beta_{m}(\Gamma_s)$.
This case is discussed in more detail in Ref.~\onlinecite{BK}.

Finally, one should note that there is no apparent reason for the
IC of $\delta$ to be restricted to values below $0.5$. In fact, the
calculations show that at high enough dopant concentration, the
effect should be very large (this is also true at $T=0$ as shown in
the next section).

\subsection{Magnetic impurity contribution to $\beta$ at $T=0$}
\label{subsec:MIT0}
Since we want to determine solely the contribution of
non-adiabaticity and
magnetic impurities to the IC, one is interested in eliminating
the presence of the BCS contribution caused by the factor
$\varphi(T/T_c)$ and the usual isotopic dependence of $T_c$.
Besides the procedure presented
in the last section there is another way to avoid $\beta_{BCS}$.
Indeed, the BCS part becomes negligibly small
($\varphi \sim 1-(T/T_c)^4$) as one lowers the temperature. At $T=0$
its contribution is zero. The penetration depth calculated by Skalski
{\it et al.} at zero temperature is given by
$\delta^{-2} = -(4\pi n e^2/mc^2)\tilde{K}(\omega=0,{\bf q}=0)$
with\cite{skalski}
\end{multicols}
\begin{eqnarray}\label{K0sl}
\tilde{K}(0,0) = -\frac{1+\bar{\Gamma}\bar{\eta}^{-3}}{\bar{\eta}}
\left[
\frac{\pi}{2}-\frac{f(\bar{\eta})}{R(\bar{\eta})}
\right]
+
\bar{\Gamma}\bar{\eta}^{-3}\left[
\frac{2}{3}\bar{\eta}-\frac{\pi}{4}\bar{\eta}+1
\right]
\end{eqnarray}
for $\bar{\Gamma},\bar{\eta}<1$ (with
$f(\bar{\eta})=\text{arcos}\bar{\eta}$) or
$\bar{\Gamma}<1$, $\bar{\eta}>1$ (with
$f(\bar{\eta})=\text{arcosh}\bar{\eta}$) and
\begin{eqnarray}\label{K0ll}
\tilde{K}(0,0) &=& - \frac{1+\bar{\Gamma}\bar{\eta}^{-3}}{\bar{\eta}}
\left\{
\frac{\pi}{2} - 2\frac{\bar{\Gamma}-1}{R(\bar{\Gamma})}
- R^{-1}(\bar{\eta})
  \left( \text{arcosh}\bar{\eta} - 2\text{artanh}{\cal R} \right)
\right\} \\
&&+ \bar{\eta}^{-3}
\left\{
   \left(\frac{2}{3}\bar{\eta}^2 + 1\right)
   \left(\bar{\Gamma} - R(\bar{\Gamma})\right)
   - \frac{1}{2}\eta R(\bar{\Gamma})\left(\frac{2}{3}\eta - 1\right)
   - \bar{\eta}\bar{\Gamma}\left( \frac{\pi}{4}
       - \frac{\bar{\Gamma}-1}{R(\bar{\Gamma})} \right)
\right\}
\nonumber
\end{eqnarray}
\begin{multicols}{2}
for $\bar{\Gamma},\bar{\eta}>1$. We have introduced the notation
$\bar{\Gamma} = \Gamma_s/\Delta$,
$\bar{\eta} = \eta\bar{\Gamma} = \Gamma_2/\Delta$,
$\eta = \Gamma_2/\Gamma_s$,
$R(x) = \sqrt{|1-x^2|}$ with $x = \bar{\Gamma},\bar{\eta}$ and
${\cal R} = [(\bar{\Gamma}-1)(\bar{\eta}-1)
/(\bar{\Gamma}+1)(\bar{\eta}+1)]^{1/2}$.
$\Delta\equiv\Delta(T=0,\Gamma_s)$ is the order parameter in the
presence of magnetic impurities.
Eqs.~(\ref{K0sl}),(\ref{K0ll}) are valid when $\Gamma_s\ll\Gamma_2$.
These two scattering amplitudes, $\Gamma_2$
and $\Gamma_s$ ($\Gamma_s = \Gamma_1 - \Gamma_2$), defined by
Abrikosov and Gor'kov\cite{AG}, describe the direct and exchange
scattering, respectively.
One can calculate the magnetic impurity contribution to the IC at
$T=0$ from Eq.~(\ref{beta}) in a straightforward way using
Eqs.~(\ref{K0sl}) and (\ref{K0ll}). The result can be written as
\begin{eqnarray}\label{icMI0}
\beta_{m}(T=0) =
-\frac{\alpha_{\Delta}}{2}\frac{K_1+K_2}{\tilde{K}(0,0)}
\end{eqnarray}
where $\alpha_{\Delta}$ is defined below, $\tilde{K}(0,0)$ is
given by Eqs.~(\ref{K0sl}), (\ref{K0ll}) and
\end{multicols}
\begin{eqnarray}\label{ICsl}
K_1 &=& - \frac{1+3\bar{\Gamma}\bar{\eta}^{-3}}{\bar{\eta}}
       \left(\frac{\pi}{2}-\frac{f(\bar{\eta})}{R(\bar{\eta})}\right)
        \pm \frac{1+\bar{\Gamma}\bar{\eta}^{-3}}{R(\bar{\eta})^2}
     \left(\frac{\bar{\eta}}{R(\bar{\eta})}f(\bar{\eta})-1\right)
\\
K_2 &=& \bar{\Gamma}\bar{\eta}^{-3}
        \left(2 - \frac{\pi}{4}\bar{\eta}\right)
\nonumber
\end{eqnarray}
for $\bar{\Gamma},\bar{\eta}<1$ (upper sign,
$f(\bar{\eta})=\text{arcos}\bar{\eta}$) or
$\bar{\Gamma}<1$, $\bar{\eta}>1$ (lower sign,
$f(\bar{\eta})=\text{arcosh}\bar{\eta}$) and
\begin{eqnarray}\label{ICll}
K_1 &=& 
- \frac{1+3\bar{\Gamma}\bar{\eta}^{-3}}{\bar{\eta}}
\left\{
\frac{\pi}{2} - 2\frac{\bar{\Gamma}-1}{R(\bar{\Gamma})}
- \frac{1}{R(\bar{\eta})}
\left(\text{arcosh}\bar{\eta} - 2\text{artanh}\cal{R}\right)
\right\}
- \frac{1+\bar{\Gamma}\bar{\eta}^{-3}}{\bar{\eta}}
\left\{
   2\frac{\bar{\Gamma}(\bar{\Gamma}-1)}{R(\bar{\Gamma})^3}
\right.
     \nonumber\\
&& \left. 
- \frac{1}{R(\bar{\eta})}
\left[
     \frac{\bar{\eta}}{R(\bar{\eta})^2}
     \left(\text{arcosh}\bar{\eta} - 2\text{artanh}\cal{R}\right)
   + \frac{{\cal R}}{1 - {\cal R}^2}
     \left(  \frac{\bar{\Gamma}^2}{R(\bar{\Gamma})^2}
           + \frac{\bar{\eta}^2}{R(\bar{\eta})^2}
     \right)
   - \frac{\bar{\eta}}{R(\bar{\eta})}
\right]
\right\} \\
K_2 &=& \frac{3}{\bar{\eta}^3}
          \left( \frac{2}{3}\bar{\eta}^2 + 1 \right)
                [\bar{\Gamma}- R(\bar{\Gamma})]
- \frac{1}{2} \eta R(\bar{\Gamma})
  \left( \frac{2}{3}\eta^2 - 1\right)
- \bar{\eta}\bar{\Gamma}
  \left( \frac{\pi}{4} - \frac{\bar{\Gamma} - 1}{R(\bar{\Gamma})}
  \right) \nonumber\\
&& + \frac{1}{\bar{\eta}^3}
\left\{
  \left[
     \frac{\bar{\Gamma}}{R(\bar{\Gamma})}
          \left( \frac{2}{3}\bar{\eta}^2 + 1\right)
          - \frac{4}{3}\bar{\eta}^2
  \right]
      [ \bar{\Gamma} - R(\bar{\Gamma}) ]
\right. \nonumber\\
&&
\left.
  + \frac{1}{2}\eta\frac{\bar{\Gamma}^2}{R(\bar{\Gamma})}
      \left( \frac{2}{3}\eta^2 - 1\right)
  + 2\bar{\eta}\bar{\Gamma}
    \left[
       \frac{\pi}{4} - \frac{\bar{\Gamma}-1}{R(\bar{\Gamma})}
          \left( 1 + \frac{\bar{\Gamma}}{2R(\bar{\Gamma})^2}\right)
    \right]
\right\}
\nonumber
\end{eqnarray}
\begin{multicols}{2}
for $\bar{\Gamma},\bar{\eta}>1$ and $\Delta T_c/T_c(\bar{\Gamma}-1)
\ll 1$. The last condition expresses the fact that the calculation
is not valid in the immediate vicinity of $\bar{\Gamma}=1$.
Eq.~(\ref{icMI0}) contains
$\alpha_{\Delta}$ which is the IC of the order parameter $\Delta$.
In strong-coupling systems, $\alpha_{\Delta}$ has to be calculated
numerically using Eliashberg's equations. In the following we
calculate the IC
in the framework of the BCS model where $\alpha_{\Delta}$ can be
calculated analytically. Indeed, from the relations
\end{multicols}
\begin{eqnarray}\label{gapAG}
\ln\left(\frac{\Delta}{\Delta_0}\right) =
\left\{ \begin{array}{l@{\quad : \quad}l}
-\frac{\textstyle \pi}{\textstyle 4}{\textstyle \bar{\Gamma}}
& \bar{\Gamma}\le 1 \\
-\ln\left[\bar{\Gamma} + R(\bar{\Gamma}) \right]
+ \frac{^{\textstyle R(\bar{\Gamma})}}{_{\textstyle 2\bar{\Gamma}}}
- \frac{^{\textstyle \bar{\Gamma}}}{_{\textstyle 2}}
\arctan R(\bar{\Gamma})^{-1}
& \bar{\Gamma} > 1
\end{array}\right.
\qquad ,
\end{eqnarray}
derived by Abrikosov and Gor'kov [$\Delta_0 = \Delta(T=0,\Gamma_s=0)$
is the order parameter in the absence of magnetic impurities] one
obtains
\begin{eqnarray}\label{ICgap}
\alpha_{\Delta} = \alpha_{\Delta_0}
\left\{ \begin{array}{l@{\quad : \quad}l}
\left( 1 - \frac{\textstyle \pi}{\textstyle 4}
{\textstyle \bar{\Gamma}} \right)^{-1}
& \bar{\Gamma}\le 1 \\
\left[
1 - \frac{^{\textstyle \bar{\Gamma}}}{_{\textstyle 2}}
\arctan R(\bar{\Gamma})^{-1} -
\frac{^{\textstyle R(\bar{\Gamma})}}{_{\textstyle 2\bar{\Gamma}}}
\right]^{-1}
& \bar{\Gamma} > 1
\end{array}\right.
\qquad .
\end{eqnarray}
\begin{multicols}{2}
In the BCS approximation one further has
$\alpha_{\Delta_0} = \alpha_0$, where the last quantity was defined
before as the IC of $T_{c0}$, that is, in the absence of magnetic
impurities. As for the calculation near $T_c$, the isotope
coefficient $\beta_{m}$
at $T=0$ is independent of any free parameter. Indeed, $\Gamma_s$ is
determined by the Abrikosov-Gor'kov equation relating $T_c$ and
$T_{c0}$, whereas $\Gamma_2$ (the direct scattering amplitude) is
determined via normal state properties of the material. Thus the
relation $\beta_{m}(T_c)$ is also universal at $T=0$.

Since the penetration depth $\delta$ does not depend explicitly
on $T_c$ at $T=0$ [see Eqs.~(\ref{K0sl}) and (\ref{K0ll})],
$\beta_{m}(T=0)$ is independent of the isotope coefficient
$\alpha$. Consequently, the contributions of magnetic impurities and
the non-adiabatic channel to the IC of $\delta$ are simply additive
at $T=0$. This has to be seen in contrast to the calculation of the
last section performed near $T_c$. In the following we neglect the
additive non-adiabatic additive contribution and show only
$\beta_{m}$.

The numerical calculation of $\beta_{m}$ from
Eqs.~(\ref{ICsl})-(\ref{ICgap}) is shown in Figs.~4a and 4b
as a function of $T_c$ (that is of the magnetic
impurity concentration) for two values of the ratio
$\eta =\Gamma_2/\Gamma_s$($>1$). Fig.~4a displays the case of a
conventional superconductor and Fig.~4b was obtained with parameters
corresponding to the situation of high-temperature superconductors.
The result shows that in analogy
with the IC of $T_c$ in the presence of magnetic impurities and
with the calculations of the last section,
$|\beta_{m}(T=0)|$ increases with increasing magnetic impurity
concentration. However, the two coefficients $\alpha$ and $\beta$
have opposite signs.

There are two main differences between the
situations of conventional and high-temperature superconductors.
On the one hand, the isotope coefficient $\beta_{m}$ at $T=0$K of
high-$T_c$
superconductors remains small over a relatively large concentration
range of magnetic impurities. The effect becomes sizeable in the
gapless regime. The increase of $|\beta_{m}(T=0)|$
is much larger at low doping for conventional superconductors. As
a consequence, on the other hand, the effect of the parameter $\eta$
is sizeable only in the case of conventional superconductors and in
the low impurity concentration regime. The calculations suggest
also that for both types of superconductors the largest effect
is observed at low $T_c$ where the superconductor is in the gapless
state ($\bar{\Gamma},\bar{\eta}>1$). The values
of the IC are then of the same order as those of the IC of $T_c$.
It would be interesting to perform experiments both
on conventional and high-temperature superconductors. 
As discussed before,  Zn- or Pr-doped YBCO are interesting
candidates since the IC of $T_c$ as a function of doping (which
is necessary in order to calculate $\beta_{tot} = \beta_{ph} +
\beta_{m} + \beta_{na}$) is known in both
cases.\cite{KBWO,franck,soerensen}

\section{ISOTOPE EFFECT OF $\delta$ IN PROXIMITY SYSTEMS}
\label{sec:prox}

In this last section we consider another physical situation,
a proximity system, in which a factor not related to lattice
dynamics, induces an IE of the penetration depth.
We have shown in Ref.~\onlinecite{KBWO} that the
presence of a normal film on a superconductor induces a change in
the IC of $T_c$. Here we show that it also induces an isotopic
shift of $\delta$.
We study a proximity system of the type $S-N$, where $S$ is
a weak-coupling superconductor and $N$ is a metal or a semiconductor
(e.g.~Pb-Ag).
Let us note $T_{c0}$ the value of the critical temperature for an
isolated $S$ film.
As is known, the proximity effect
leads to a critical temperature $T_c$ that differs from
$T_{c0}$ (see, e.g., Refs.~\onlinecite{KBWO,kresin3}). The proximity
effect also affects the shielding of a weak
magnetic field. The most dramatic effect of the normal
layer on the penetration depth is seen in the low temperature regime
($T/T_c \le 0.3$). Although the penetration depth of a pure
conventional superconductor is only weakly temperature dependent in
this regime ($\delta \sim \sqrt{1-T^4/T_c^4}$ ) the presence of the
normal layer induces a temperature dependence through the proximity
effect.

Consider a proximity system $S-N$ and assume that
$\delta < L_N \ll \xi_N$ (which is certainly
satisfied in the low-temperature regime since $\xi_N = \hbar v_F/
2\pi T$; see, e.g., Ref.~\onlinecite{kresin4}), where $L_N$
and $\xi_N$ are the thickness and the coherence length of the normal
film, respectively (cf.~Ref.~\onlinecite{KBWO} and references
therein).
In this case,
the penetration depth $\delta$ has been calculated by one of the
authors\cite{kresin4}:
\begin{eqnarray}\label{prox}
\delta^{-3} = a_N \Phi \qquad ,
\end{eqnarray}
where $a_N$ is a constant depending only on the material
properties of the normal film (it is independent of the ionic mass)
and
\begin{eqnarray}\label{phi}
\Phi = \pi T \sum_{n \ge 0} \frac{1}{x_n^2p_n^2 + 1}
\, , \qquad p_n = 1+ \varepsilon t \sqrt{x_n^2 + 1} \qquad .
\end{eqnarray}
$x_n = \omega_n/\epsilon_S(T)$ with $\omega_n = (2n+1)\pi T$ (the
Mastubara frequencies) and $\epsilon_S(T)$ is the superconducting
energy gap of $S$. In the weak-coupling limit considered here,
$\epsilon_S(0) = \varepsilon\pi T_{c0}$ with $\varepsilon \simeq
0.56$. The dimensionless parameter $t = \ell/S_0$
with $\ell = L_N/L_0$ and $S_0 = \Gamma_0/\pi T_{c0}$
($\Gamma_0 \sim 1/L_0$ is the McMillan parameter\cite{millan}).
$L_N$ and $L_0$ are the thickness of the normal film and some
arbitrary thickness, respectively (in the following we take
$L_0 = L_S$, the thickness of the superconducting film).

Since $\delta$ depends non-linearly on $T_{c0}$ (through the
proximity parameter $t$), the penetration depth will display
an isotope shift due to the proximity effect.
From Eq.~(\ref{prox}) one can calculate the isotope coefficient
$\beta_{prox}$ [as
defined by Eq.~(\ref{beta})] of the penetration depth due to the
proximity effect:
\begin{eqnarray}\label{betaprox}
\beta_{prox} = - \frac{2\alpha_0}{3\Phi}
\sum_{n>0} \frac{x_n^2p_n^2}{x_n^2p_n^2 + 1}
\left( 1 - \frac{\varepsilon t}{p_n\sqrt{x_n^2 + 1}} \right) \, ,
\end{eqnarray}
where $\alpha_0$ is the IC of $T_{c0}$ for the superconducting film S
alone.

This result has several interesting properties. First, it
turns out that the isotope
coefficients $\beta_{prox}$ and $\alpha_0$ have generally opposite
signs (see below). Secondly, the IC depends on the proximity
parameter $t$, that is on the thickness ratio $\ell = L_N/L_S$ of
the normal and superconducting films and on the MacMillan tunneling
parameter $\Gamma_0$.

The dependency on the ratio $\ell$ has already
been mentioned in connection with the IE of $T_c$\cite{KBWO,BK}.
Fig.~5 shows the isotope coefficient $\beta_{prox}$ as a function of
the parameter $t$ for fixed temperatures. $|\beta_{prox}|$ decreases
with increasing $t$. There are two ways to change the value of
$t\sim \ell/S_0$.
One of them is an increase of the parameter $\ell= L_N/L_S$ (e.g.~an
increase of the thickness of the normal film $L_N$). Note
that there is a lower
bound $\ell_{min}$ to the value of $\ell$, since it was assumed that
$\delta < L_N$ to obtain Eq.~(\ref{prox})\cite{kresin4}. The IC
$|\beta_{prox}|$ decreases with
increasing thickness ratio (Fig.~5 then corresponds
to the values $S_0 = 0.2$ for $\ell \in ]\ell_{min},1]$). However,
the higher the
temperature the smaller the decrease of $\beta_{prox}$ which then
rapidly reaches saturation with increasing thickness.

Instead of changing the thickness ratio $\ell$ one can vary the
parameter $\Gamma_0$. One possibility is to use different films,
another is to modify the quality of the $S-N$ interface.
Fig.~5 then implies that the $|\beta_{prox}|$
decreases with decreasing tunneling parameter $\Gamma_0$.
It would be interesting to perform an experiment where the isotope
effect of $T_c$ and $\delta$ are determined for a proximity system
with different values of the thickness ratio $\ell$.

Note that the result (\ref{betaprox}) has another interesting feature
; one can see that the isotopic shift of the
penetration depth is {\it temperature dependent}. Fig.~6 shows the
temperature dependence of $\beta_{prox}$ for two sets of parameters.
The two upper curves are obtained for $S_0 = 0.2$ and two thickness
ratios $\ell=0.5,\, 1$; the two lower curves are for $S_0=1$ and
$\ell= 0.5,\, 1$. From Fig.~6 one concludes that
$|\beta_{prox}|$ increases with increasing temperature. Furthermore,
a weak tunneling between the superconducting and the normal film
(smaller $S_0$) induces a strong temperature dependence. Finally, the
temperature range over which the IC remains unchanged is broader for
a larger than a smaller tunneling parameter.

Therefore, the presence of a normal layer on a
superconductor induces an isotope shift of the penetration depth.
Moreover, in a proximity system the IC depends on the thickness
ratio and the temperature. The values of the isotope
coefficient are such that they are measurable in conventional and
high-temperature superconductors.

\section{CONCLUSION}\label{sec:concl}

We have shown that the value of such a fundamental parameter as the
penetration depth $\delta$ depends on isotopic substitution. This
unconventional dependency is caused by number of factors not related
to the pairing mechanism.

At first, we focused on the non-adiabatic IE. This effect, manifested
in the dependence of the charge-carrier concentration on the ionic
masses for doped systems such as cuprates\cite{KW,KBWO}, leads to a
noticeable
isotopic shift of the penetration depth $\delta$.
For this case, we established a general relation between the isotope
coefficient of $T_c$ and of $\delta$ for London superconductors
(see Eq.~(\ref{ICTcNA})). The isotope coefficient of $\delta$ is a
constant independent of doping or temperature; we
determined its value for oxygen depleted and Pr substituted
YBa$_2$Cu$_3$O$_{7-\delta}$ as well as for La$_{2-x}$Sr$_x$CuO$_4$.
Note that similar effects are expected in manganites\cite{KW2} where
the charge-transfer processes also involve the non-adiabatic oxygen
ions.

The presence of magnetic impurities also leads to a non-trivial
isotope effect for $\delta$. We studied this effect for temperatures
near $T_c$ and at $T=0$. It turns out that the
coefficient obeys a universal law; it is determined by
the experimental values of $T_c$ in the absence and presence of
magnetic impurities [Eqs.~(\ref{icMI}) and (\ref{icMI0})] as well as
the isotope coefficient of $T_c$ in the absence of magnetic
impurities. The combined
effect of magnetic impurities and non-adiabaticity was also
discussed. Finally, we showed that the penetration depth for
proximity systems also display an isotopic shift. It is interesting
that in the case of a proximity system or in the presence of
magnetic impurities the IC for $\delta$ is temperature
dependent.

All these effects should be measurable in conventional and in
high-temperature superconductors and several experiments were
proposed. In the case of high-$T_c$ materials it would be of interest
to study oxygen depleted YBa$_2$Cu$_3$O$_{7-\delta}$ as well as the
Pr- and Zn-substituted systems, since the IE of $T_c$ has
been successfully described in these materials
(see Refs.~\onlinecite{KW,KBWO,BK}). The case of
Zn-substituted YBa$_2$Cu$_3$O$_{6+x}$ is here of special interest,
since experiments show that the depression of $T_c$ is mainly due to
the change in magnetic impurity concentration and our theory is free
of any adjustable parameter in this case.

 Note, finally, that not only $T_c$ and $\delta$ exhibit the
unconventional isotope effects studied here and in
Refs.~\onlinecite{KW,KBWO,BK},
but any quantity that depends on the charge carrier density of the
superconducting condensate or/and on $T_c$ (e.g.~the specific
heat).\\

A.B.~is grateful to the Swiss National Science Foundation and the
Naval Research Laboratory for financial support. The work of
V.Z.K.~is supported by the U.S.~Office of Naval Research under
contract No.~N00014-96-F0006.


\end{multicols}

\newpage
{\bf Figure captions:}
\begin{description}

\item{Fig.~1:} Temperature dependency (near $T_c$) of the isotope
coeff.~$\beta$ a) in the presence of magnetic impurities only,
with $T_c/T_{c0} \simeq 0.35$ (solid), $0.5$ (dashed line), and
$\alpha_0 = 0.025$ (corresponds to
YBa$_2$(Cu$_{1-x}$Zn$_x$)$_3$O$_{7-\delta}$);
b) in the presence of magnetic impurities and non-adiabaticity for
$T_c/T_{c0} \simeq 0.5$ (dash-dotted), $0.6$ (dotted line).
$\alpha_0 = 0.025$, $\gamma_s = 0.16$ (corresponds to
Y$_{1-x}$Pr$_x$Ba$_2$Cu$_3$O$_{7-\delta}$, see text).
\item{Fig.~2:} Isotope coefficient $\tilde{\beta}_{m}$ near $T_c$ as a function of
magnetic impurity concentration for $T/T_c = 0.75$ (solid line),
$0.85$ (dashed) and $0.95$ (dotted). $\alpha_0 = 0.025$ (corresponds
to the situation of YBa$_2$(Cu$_{1-x}$Zn$_x$)$_3$O$_{7-\delta}$).
\item{Fig.~3a:} $\tilde{\beta}_{m+na}$ as a function of magnetic impurity
concentration in the presence of non-adiabaticity for
$T/T_c = 0.75$ (solid line), $0.85$ (dashed) $0.95$ (dotted).
$\alpha_0 = 0.025$, $\gamma = 0.16$ and $\tilde{\Gamma} = 123$K
(parameters for Y$_{1-x}$Pr$_x$Ba$_2$Cu$_3$O$_{7-\delta}$).
\item{Fig.~3b:} Low doping regime of Fig.~3a. In addition, the case without
non-adiabaticity is added for comparison (long dashed line,
corresponding to YBa$_2$(Cu$_{1-x}$Zn$_x$)$_3$O$_{7-\delta}$ with
$T/T_c = 0.9$).
\item{Fig.~4a:} Isotope coeff.~$\beta_{m}$ for conventional superconductors
at $T=0$ as a function of magnetic impurity concentration for
$\Gamma_2/\Gamma_s = 10$ (solid line) and $50$ (dashed).
$\alpha_0 = 0.3$.
\item{Fig.~4b:} Isotope coeff.~$\beta_{m}$ for high-$T_c$ superconductors
at $T=0$ as a function of magnetic impurity concentration for
$\Gamma_2/\Gamma_s = 10$ (solid line) and $50$ (dashed)
$\alpha_0 = 0.025$.
\item{Fig.~5:} Isotope coeff.~$\beta_{prox}$ for a proximity system as a
function of $t$ for $T/T_c = 0.1$ (solid line), $0.2$ (dashed) and
$0.3$ (dotted). $\alpha_0 = 0.5$.
\item{Fig.~6:} Isotope coeff.~$\beta_{prox}$ for a proximity system as a
function of $T/T_c$ for $S_0=0.2$: $l=1$ (dashed) $l=0.5$ (solid) and
$S_0=1$: $l=1$ (small dots) $l=0.5$ (dotted). $\alpha_0 = 0.5$.
\end{description}

\newpage
\thispagestyle{empty}
\unitlength1cm
\begin{figure}[h]
\begin{center}
\input{fig1AB.ps}
\end{center}
\vspace*{2cm}
\end{figure}
\centerline{Figure 1}
\vspace*{5cm}
\centerline{A.~Bill et al.}

\newpage
\thispagestyle{empty}
\unitlength1cm
\begin{figure}[h]
\begin{center}
\input{fig2AB.ps}
\end{center}
\vspace*{2cm}
\end{figure}
\centerline{Figure 2}
\vspace*{5cm}
\centerline{A.~Bill et al.}

\newpage
\thispagestyle{empty}
\unitlength1cm
\begin{figure}[h]
\begin{center}
\input{fig3AB.ps}
\end{center}
\vspace*{2cm}
\end{figure}
\centerline{Figure 3a}
\vspace*{5cm}
\centerline{A.~Bill et al.}

\newpage
\thispagestyle{empty}
\unitlength1cm
\begin{figure}[h]
\begin{center}
\input{fig4AB.ps}
\end{center}
\vspace*{2cm}
\end{figure}
\centerline{Figure 3b}
\vspace*{5cm}
\centerline{A.~Bill et al.}

\newpage
\thispagestyle{empty}
\unitlength1cm
\begin{figure}[h]
\begin{center}
\input{fig5AB.ps}
\end{center}
\vspace*{2cm}
\end{figure}
\centerline{Figure 4a}
\vspace*{5cm}
\centerline{A.~Bill et al.}

\newpage
\thispagestyle{empty}
\unitlength1cm
\begin{figure}[h]
\begin{center}
\input{fig6AB.ps}
\end{center}
\vspace*{2cm}
\end{figure}
\centerline{Figure 4b}
\vspace*{5cm}
\centerline{A.~Bill et al.}

\newpage
\thispagestyle{empty}
\unitlength1cm
\begin{figure}[h]
\begin{center}
\input{fig7AB.ps}
\end{center}
\vspace*{2cm}
\end{figure}
\centerline{Figure 5}
\vspace*{5cm}
\centerline{A.~Bill et al.}

\newpage
\thispagestyle{empty}
\unitlength1cm
\begin{figure}[h]
\begin{center}
\input{fig8AB.ps}
\end{center}
\vspace*{2cm}
\end{figure}
\centerline{Figure 6}
\vspace*{5cm}
\centerline{A.~Bill et al.}


\begin{references}{}
\bibitem{KW}
V.Z.~Kresin, and S.A.~Wolf, Phys.~Rev.~B {\bf 49}, 3652 (1994);
and in {\it Anharmonic Properties of High-$T_c$ Cuprates},
p.~18, D.~Mihailovic, G.~Ruani, E.~Kaldis, K.A.~M\"uller, Eds., World
Scientific (1995).
\bibitem{KBWO}
V.Z.~Kresin, A.~Bill, S.A.~Wolf, and Yu.N.~Ovchinnikov, Phys.~Rev.~B
{\bf 56}, 107 (1997).
\bibitem{BK}
A.~Bill, V.Z.~Kresin, and S.A.~Wolf, Z.~Phys.~B, in press;
A.~Bill, and V.Z.~Kresin, Proceedings of the International Symposium
on {\it Electrons and Vibrations in Solids and Finite Systems},
Berlin, August 1996, U.~Scherz and H.-J.~Schulz eds.,
Z.~Phys.~Chem.~{\bf 201}, 271 (1997).
\bibitem{franck}
J.P.~Franck, J.~Jung, M.A-K.~Mohamed, S.~Gygax, and G.I.~Sproule,
Phys.~Rev.~B {\bf 44}, 5318 (1991).
\bibitem{soerensen}
G.~Soerensen, and S.~Gygax , Phys.~Rev.~B {\bf 51}, 11848 (1995).
\bibitem{zech}
D.~Zech, K.~Conder, H.~Keller, E.~Kaldis, and K.A.~M\"uller,
Physica B {\bf 219\&220}, 136 (1996).
\bibitem{morris}
H.J.~Bornemann and D.E.~Morris, Phys.~Rev.~B {\bf 44}, 5322 (1991).
\bibitem{zhao1}
G.-M.~Zhao and D.E.~Morris, Phys.~Rev.~B {\bf 51}, 16487 (1995).
\bibitem{zhao2}
G.-M.~Zhao, K.K.~Singh, A.P.B.~Sinha, and D.E.~Morris,
Phys.~Rev.~B {\bf 52}, 6840 (1995); G.-M.~Zhao, M.B.~Hunt, H.~Keller,
and K.A.~M\"uller, Nature {\bf 385}, 236 (1997); G.~Deutscher,
Proceedings of the International School on {\it The Gap
Symmetry and Fluctuations in High-$T_c$ Superconductors},
Carg\a`ese (1997).
\bibitem{KW2}
V.Z.~Kresin, and S.A.~Wolf, Phil.~Mag.~{\bf 76}, 241 (1997).
\bibitem{mustre}
J.~Mustre de Leon, S.D.~Conradson, I.~Batistic, and A.R.~Bishop,
Phys.~Rev.~Lett.~{\bf 65}, 1675 (1990);
L.V.~Gasparov, V.D.~Kulakovskii, V.B.~Timofeev, and E.Ya.~Sherman,
J.~Supercond.~{\bf 8}, 27 (1995); G.~Ruani, P.~Guptasarma, and C.~Taliani,
Solid State Comm.~{\bf 96}, 653 (1995); A.~Jezowski, J.~Klamut, and
B.~Dabrowski, Phys.~Rev.~B {\bf 52}, 7030 (1995).
\bibitem{sharma}
R.P.~Sharma, T.~Venkatesan, Z.H.~Zhang, J.R.~Liu, R.~Chu, and W.K.~Chu,
Phys.~Rev.~Lett.~{\bf77}, 4624 (1997);
D.~Haskel, E.A.~Stern, D.G.~Hinks, A.W.~Mitchell, and J.D.~Joergensen,
Phys.~Rev.~B {\bf 56}, 521 (1997).
\bibitem{johnson}
K.H.~Johnston, D.P.~Clougherty, and M.E.~McHenry,
in {\it Novel Superconductivity}, p.~563,
S.~Wolf and V.Z.~Kresin, Eds., Plenum, NY (1986).
\bibitem{omalley}
T.F.~O'Malley, Phys.~Rev.~{\bf 162}, 98 (1967).
\bibitem{lynton}
E.A.~Lynton, {\it Superconductivity}, p.~37, Methuen Eds.,
1969.
\bibitem{crawford}
M.K.~Crawford, M.N.~Kunchur, W.E.~Farneth, E.M.~McCarron III,
and S.J.~Poon, Phys.~Rev.~{\bf 41}, 282 (1990)
\bibitem{dabrowski}
M.I.~Salkola, V.J.~Emery, and S.A.~Kivelson,
Phys.~Rev.~Lett.~{\bf 77}, 155 (1996);
B.~Dabrowski, Z.~Wang, K.~Rogacki, J.D.~Jorgensen, R.L.~Hitterman,
J.L.~Wagner, B.A.~Hunter, P.G.~Radaelli, and D.G.~Hinks,
Phys.~Rev.~Lett.~{\bf 76}, 1348 (1996);
Physica C {\bf 217}, 455 (1993);
N.~Yamada and M.~Ido, Physica C {\bf 203}, 240 (1992);
P.G.~Radaelli, D.G.~Hinks, A.W.~Mitchell, B.A.~Hunter, J.L.~Wagner,
B.~Dabrowski, K.G.~Vandervoort, H.K.~Viswanathan, and J.D.~Jorgensen,
Phys.~Rev.~B {\bf 49}, 4163 (1994);
B.~B\"uchner, M.~Breuer, A.~Freimuth, and A.P.~Kampf,
Phys.~Rev.~Lett.~{\bf 73}, 1841 (1994);
M.R.~Norman, G.J.~McMullan, D.L.~Novikov, and A.J.~Freeman,
Phys.~Rev.~B {\bf 48}, 9935 (1993).
\bibitem{AG}
A.~Abrikosov, and L.~Gor'kov, Sov.~Phys.~JETP {\bf 12}, 1243 (1961).
\bibitem{skalski}
S.~Skalski, O.~Betbeder-Matibet, and P.R.~Weiss,
Phys.~Rev.~{\bf 136}, 1500 (1963).
\bibitem{carbotte}
J.P.~Carbotte, M.~Greeson, and A.~Perez-Gonzalez, Phys.~Rev.~Lett.~{\bf 66},
1789 (1991);
S.P.~Singh, R.K.~Pandey, and P.~Singh, J.~Supercond.~{\bf 9}, 269 (1996).
\bibitem{zagoulev}
H.~Alloul, P.~Mendels, H.~Casalta, J.F.~Marucco, and J.~Arabski,
Phys.~Rev.~Lett.~{\bf 67}, 3140 (1991);
T.~Miyatake, K.~Yamaguchi, T.~Takata, N.~Koshizuka, and S.~Tanaka,
Phys.~Rev.~B {\bf 44}, 10139 (1991);
A.V.~Mahajan, H.~Alloul, G.~Collin, and J.F.~Marucco,
Phys.~Rev.~Lett.~{\bf 72}, 3100 (1994); S.~Zagoulaev, P.~Monod, and
J.~Jegoudez, Phys.~Rev.~B {\bf 52}, 10474 (1995);
Physica C {\bf 259}, 271 (1996);
M.B.~Maple, C.C.~Almasan, C.L.~Seaman, S.H.~Han, K.~Yoshiara, M.~Buchgeister,
L.M.~Paulius, B.W.~Lee, D.A.~Gajewski, R.F.~Jardim, C.R.~Fincher,
G.B.~Blanchet,and R.P.~Guertin, J.~Supercond.~{\bf 7}, 97 (1994).
\bibitem{furrer}
J.~Mesot, and A.~Furrer, J.~Supercond.~(in press); A.~Furrer,
J.~Mesot, P.~Allenspach, U.~Staub, F.~Fauth, and M.~Guillaume, in
{\it Phase Separation in Cuprate Superconductors}, p.~101,
E.~Sigmund and K.A.~M\"uller Eds., Springer-Verlag, Berlin (1994);
O.N.~Bakharev, A.V.~Dooglav, A.V.~Egorov, O.B.~Marvin, V.V.~Naletov,
M.A.~Teplov, A.G.~Volodin, and D.~Wagener, {\it ibid.}, p.~257;
Phys.~Rev.~B {\bf 55}, 11839 (1997); J.~Supercond.~{\bf 8}, 413
(1995); N.~Phillips, R.~Fisher, and
J.~Gordon, in {\it Progress in Low-Temperature Physics}
{\bf 13}, D.~Brewer Ed.~(North-Holland, Amsterdam, 1992), p.~267;
J.M.~Wade, J.W.~Loram, K.A.~Mirza, J.R.~Cooper, and J.L.~Tallon,
J.~Supercond.~{\bf 7}, 261 (1994); A.~Junod, in
{\it Studies of High Temperature Superconductors}, A.~Narlikar
Ed.~(Nova Science, Commack, NY, 1996).
\bibitem{kresin3}
V.Z.~Kresin, Phys.~Rev.~B {\bf 25}, 157 (1982).
\bibitem{kresin4}
V.Z.~Kresin, Phys.~Rev.~B {\bf 32}, 145 (1985).
\bibitem{millan}
W.~McMillan, Phys.~Rev.~{\bf 175}, 537 (1968).
\end{references}
\end{document}